\documentclass[conference,a4paper]{IEEEtran}
\IEEEoverridecommandlockouts
\pdfoutput=1
\usepackage{cite}
\usepackage{amsmath,amssymb,amsfonts}
\usepackage{graphicx}
\usepackage{algpseudocode}
\usepackage{textcomp}
\usepackage[table,xcdraw]{xcolor}
\def\BibTeX{{\rm B\kern-.05em{\sc i\kern-.025em b}\kern-.08em
    T\kern-.1667em\lower.7ex\hbox{E}\kern-.125emX}}
\usepackage{svg}
\usepackage[citecolor=black,
            colorlinks=true,
            linkcolor=black,
            urlcolor  = black,
            pdfpagelabels={true},
            pdftitle={MetricGAN+/-: Increasing Robustness of\\ Noise Reduction on Unseen Data},
            pdfauthor={George Close, Thomas Hain, and Stefan Goetze},
            pdfkeywords={speech enhancement, speech quality metrics, neural networks, GAN},
            pdfsubject={speech enhancement, speech quality metrics, neural networks, GAN, metric prediction}]
            {hyperref}
\usepackage{acronym} 
\acrodef{ASR} {Automatic Speech Recognition}
\acrodef{DFT} {discrete Fourier transform}
\acrodef{GAN} {Generative Adversarial Network}
\acrodef{h2h} {human-to-human}
\acrodef{h2m} {human-to-machine}
\acrodef{MOS} {Mean Option Score}
\acrodef{NN}  {neural network}
\acrodef{PESQ}{Perceptual Evaluation of Speech Quality}
\acrodef{SE}  {speech enhancement}
\acrodef{SNR} {Signal to Noise Ratio}
\acrodef{STFT}{Short Time Fourier Transform}
\acrodef{STOI}{Short-Time Objective Intelligibility}
\acrodef{VAD} {Voice Activity Detector}
\acrodef{WER} {Word Error Rate}
\acrodef{OLA}{Overlap-Add}
\acrodef{MSE}{mean squared error}

\begin{document}

\title{MetricGAN+/-: Increasing Robustness of\\ Noise Reduction on Unseen Data
\thanks{This work was supported by the Centre for Doctoral Training in Speech and Language Technologies (SLT) and their Applications funded by UK Research and Innovation [grant number EP/S023062/1]. This work was also funded in part by TOSHIBA Cambridge Research Laboratory.}}


\author{\IEEEauthorblockN{George Close, Thomas Hain, and Stefan Goetze\\}
\IEEEauthorblockA{
\textit{Dept.~of Computer Science, {The} University of Sheffield,} 
Sheffield, United Kingdom \\
\{glclose1, t.hain, s.goetze\}@sheffield.ac.uk\vspace*{-0.25cm}
}
}

\maketitle

\begin{abstract}
Training of speech enhancement systems often does not incorporate knowledge
of human perception and thus can lead to unnatural sounding results. Incorporating psychoacoustically motivated speech perception metrics as part of model training via a predictor network has recently gained interest. However, the performance of  such predictors is limited by the distribution of metric scores that appear in the training data. \\
In this work, we propose MetricGAN+/- (an extension of MetricGAN+, one such metric-motivated system) which introduces an additional network - a ``de-generator'' 
 to improve the robustness of the prediction network (and by extension of the generator) by ensuring observation of a wider range of metric scores in training. Experimental results on the VoiceBank-DEMAND dataset show relative improvement in PESQ score of $3.8\%$ ($3.05$ vs. $3.22$ PESQ score), as well as better generalisation to  unseen noise and speech signals.  

\end{abstract}


\section{Introduction}

\Ac{SE} has been an active research topic for decades now, given its myriad applications in 
\ac{h2h} 
communication in video or voice calls as well as in 
\ac{h2m} 
communication in home, industry and mobile device assistant products \cite{FFASRHaebUmbach,XMM+15}.
Use of \ac{NN} systems to perform speech enhancement has shown great success in recent years \cite{barker_asru2015, reddy2021interspeech,RGH22_AttTASNET,dolcoMVDRenhance,Moritz2017}. 
{Training of \acp{NN}} 
{for} 
speech enhancement 
{requires selection of}
an objective function 
appropriate for 
{the} task. Direct comparison between `clean' audio and the output of a neural network given an artificially corrupted version of that audio has been found to be only weakly correlated  with objective measures (metrics) of intelligibility, quality and performance for both forms of speech communication \cite{GWK+14,bagchi2018spectral,Chai2018AcousticsguidedE}. 
A recent publication \cite{stoi_loss}  proposed a loss function that corresponds to one of these psychoacoustically motivated metrics. However, such objective functions must be carefully designed as many objective measures contain calculations that are non-differentiable. Several systems circumvent this limitation via use of an additional model that mimics the behaviour of the metric \cite{fu2018qualitynet,PESQNet,PESQNet2}, with this network being used as a surrogate of the metric used as an objective function in training of the speech enhancement model. 
The baseline system that this work builds upon is one such system, MetricGAN+ \cite{fu21_interspeech} (itself an extension of previous work MetricGAN \cite{FuLTL19}).
Two popular objective measures the \ac{PESQ} \cite{pesq} and the \ac{STOI} \cite{stoi} for speech quality and intelligibility respectively are used. Both measures account for human perception and are often highly correlated with \ac{MOS} of human evaluators \cite{stoi_loss,GWK+14,reddy2021dnsmos}. 
The computation of \ac{STOI} is relatively simple, and a version of it suitable for use as an objective function is detailed in  \cite{stoi_loss}. Calculation of PESQ is  more complex, and thus cannot be formulated in a differentiable way to be used as objective functions.
To handle this problem, a secondary "discriminator" network is introduced that, given a representation of the reference and the degraded signal, predicts the metric score corresponding to those two signals.  Such a metric prediction network is sometimes referred to as a QualityNet~\cite{fu2018qualitynet}.
The output of this discriminator network is then used to train the speech enhancement (generator) network. The two networks are trained in a \ac{GAN} style strategy. In this work we  introduce a further network, a 'de-generator' which attempts to produce outputs with a set lower metric score, aiming to improve the ability of the discriminator to predict the metric on a more complete range of metric scores. 
\\
The 
{remainder} of this paper is structured as follows: Section~\ref{sect:metricgan+} presents the baseline system, its model structure and training setup. The proposed extension is introduced in 
{Section}~\ref{mg-}, followed by a comparison to the baseline in Section \ref{sect:Results} and a brief conclusion in Section~\ref{sect: Conclusion}.
%
%
%
\section{Baseline System - MetricGAN+}\label{sect:metricgan+}
%
The MetricGAN+ framework \cite{fu21_interspeech} consists of two networks: a speech enhancement model $\mathcal{G}$, which aims to remove the undesired signal parts, i.e the noise $v[n]$ from a noisy signal
\begin{equation}
    x[n] = s[n] + v[n]
\end{equation}
to produce an estimate of a clean signal $s[n]$, denoted in the following by $\hat{s}[n]$ 
and a metric discriminator (more correctly an evaluator) $\mathcal{D}$, which predicts {the} possibly psychoacoustically motivated performance metrics(s) {providing a target} to optimise the signal enhancement.
%
\subsubsection{Input Features} 
%
\begin{figure}[!ht]
\centering
    \resizebox{0.75\columnwidth}{!}{%
    \def\svgwidth{1\columnwidth} 
 	\graphicspath{{figures/}}     
 	\resizebox{\columnwidth}{!}{
    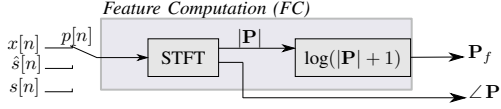 
    }
    }
\caption{\label{fig:features}Illustration of feature computation (FC), $p[n]$ can be any signal $x[n],\hat{s}[n],s[n]$
}
\end{figure}
 Features are calculated from  different time domain signals denoted here by $p[n]$  and which is a placeholder for the signals ${x[n],s[n],\hat{s}[n]}$ for the time index $n$ which we will omit in the following. First a spectral magnitude $P_{k,\ell}$ for frequency $k$ and frame $\ell$ is calculated  of the time domain audio signal $p$ using the \ac{STFT}, followed by transformation to the feature space by adding $1$ to and taking the logarithm of each element to give the feature representation $\mathbf{P}_f$ as shown in Fig.~\ref{fig:features}.
The phase of the spectral bins  $\angle p_{k,\ell}$ will be used later to resythesize the time domain signal using the \ac{OLA} method. 
%
\subsubsection{{Generator Network for Signal Enhancement}}
\begin{figure}[!ht]
    \centering
    \def\svgwidth{1.8\columnwidth} 
 	\graphicspath{{figures/}}     
 	\resizebox{\columnwidth}{!}{
    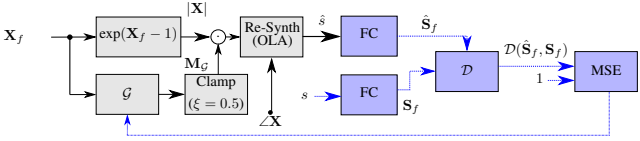 
    }
    \caption{\textcolor{blue}{Training} and \textcolor{gray}{inference} of MetricGAN+ Generator.
    }
    \label{fig:gen_train}
\end{figure}

%
Fig.~\ref{fig:gen_train} shows the training of $\mathcal{G}$.  The dotted \textcolor{blue}{blue} arrows and processes show the objective function and loss calculation back-propagated to the model. 
In order to obtain the enhanced signal $\hat{s}$ from  the noisy features
$\mathbf{X}_f$ in the generator $\mathcal{G}$'s training and inference,   the transform is reversed by subtracting $1$ from each element and taking the exponential of each element in the feature representation. 
The output of $\mathcal{G}$ is a time-frequency (T-F) mask matrix $\mathbf{M}_\mathcal{G}$ 
, which is then multiplied with the noisy magnitude spectogram $|\mathbf{X}|$  to 
result in the enhanced signal matrix
$|\mathbf{\hat{S}}|$. The enhanced time domain audio signal 
$\hat{s}[n]$ is calculated using \ac{OLA} resythesis.
Note that  each element in $\mathbf{M}_\mathcal{G}$ is `clamped' in order to reduce residual musical tones caused by the mask, i.e.~it is limited to element wise values $\xi \leq \mathbf{M}_\mathcal{G} \leq 1 $.
The objective function of the speech enhancement network $\mathcal{G}$ is dependent entirely on the metric score of its output $\hat{s}$ (in its feature space representation $\mathbf{\hat{S}}_f$) as predicted by discriminator $\mathcal{D}$
\begin{equation}
\label{eq:2}
L_{\mathcal{G},\mathrm{MG+}} = \mathbb{E}[(\mathcal{D}(\mathbf{\hat{S}}_f,\mathbf{S}_f) -1)^2]
\end{equation}
where $1$ represents a `perfect' score in the normalised metric $Q'(\cdot)$.

\subsubsection{{Discriminator Network for Metric Prediction}}

\begin{figure}[!ht]
\centering
   \def\svgwidth{1.7\columnwidth} 
 	\graphicspath{{figures/}}     
 	\resizebox{\columnwidth}{!}{
    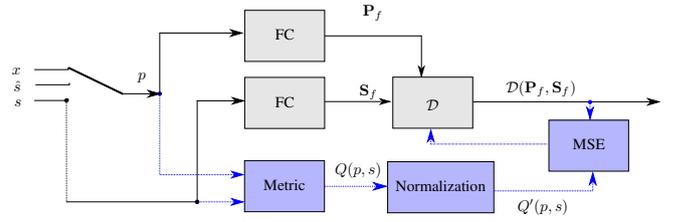 
    }
\caption{\textcolor{blue}{Training} and \textcolor{gray}{inference} of MetricGAN+ Discriminator.}\label{fig:discrim_train}
\end{figure}

The discriminator $\mathcal{D}$ is trained to reproduce the normalised target metric $Q'(\cdot)$ minimising the distance from its output and the `true' normalised metric score used as its objective function, as visualised in  Fig.~\ref{fig:discrim_train}. Arrows and processes marked \textcolor{blue}{blue} denote those which occur only during training.
The loss of the discriminator comprises three \ac{MSE} terms depending on the clean reference signal $s$, or $\mathbf{S}_f$, the degraded noisy signal $x$, or $\mathbf{X}_f$, and the enhanced signal $\hat{s}$, or $\mathbf{\hat{S}}_f$. More specifically, its objective function is given as:
%
\begin{multline}
\label{eq:1}
\hspace*{-1.5ex}
L_{D,MG+} = \mathbb{E}[(\mathcal{D}(\mathbf{S}_f,\mathbf{S}_f) -1)^2+
(\mathcal{D}(\mathbf{\hat{S}_f},\mathbf{S}_f) - Q'(\hat{s},s))^2 +\\
(\mathcal{D}(\mathbf{X}_f,\mathbf{S}_f)- Q'(x,s))^2]
\end{multline}

%
%
The $1$ in the first term of (\ref{eq:1}) represents the fact that $Q'(s,s) =1$. In the second term, the scores of signals enhanced by $\mathcal{G}$, $\hat{s}$ are considered and compared to the ground truth score for the enhanced signal. In the final term, the scores of noisy signals $x$ are considered and compared to the true score for the noisy signal. Note that in the case of the metrics investigated in this work the input to the function that defines the metric are the time domain signals $x$, $\hat{s}$ and $s$, but this may not always be the case.

\subsubsection{MetricGAN+ Training}

Each epoch of training consists of four steps, the first three representing the training of $\mathcal{D}$ and the final step the training of $\mathcal{G}$. At the start of each epoch, $I$ audio segments are randomly picked out from the training set. Firstly $\mathcal{D}$ is trained as given in (\ref{eq:1}) on these $I$ random  audio segments. The audio segments are time domain signals of varying length. Then, in the second step, $\mathcal{D}$ is trained using a 'replay buffer' where saved enhanced outputs of the generator $\mathcal{G}$ from past epochs are used to train $\mathcal{D}$. The size of this replay buffer is decided by a `\texttt{history\_portion}' $H$ hyper-parameter, which corresponds to the replay buffer growing by a set percentage of the audio segments observed each epoch. This is done to prevent $\mathcal{D}$ from `forgetting' too much about the behaviour of $Q'(\cdot)$ on previously enhanced speech.\\ Then the first step is repeated with $\mathcal{D}$ again being trained using the $t$ random samples. Finally, $\mathcal{G}$ is trained also using these $t$ samples as in (\ref{eq:2}). During training of $\mathcal{D}$, $\mathcal{G}$ is `frozen' and its parameters are not updated; the opposite is true during $\mathcal{G}$'s training. Note that samples are added to the replay buffer during the first step of $\mathcal{D}$'s training, meaning that 20\% of the `current' epoch data are always present in the replay buffer. As $\mathcal{D}$ is trained before $\mathcal{G}$, the $\hat{s}$ in (\ref{eq:1}) actually represents the output of the previous epoch's $\mathcal{G}$.

\subsubsection{Discriminator Model Structure}
%
The discriminator $\mathcal{D}$'s structure is a Convolutional Neural Network (CNN) with four 2D convolutional layers with 15 filters of a kernel size of $(5,5)$. To account for the variable length of input data, a global 2-D average pooling layer is placed immediately after the input, fixing the feature representation at $15$ dimensions. After the convolutional layers, a mean is taken over the $2$nd and $3$rd dimensions, and this representation is fed into three sequential linear layers, with $50$, $10$ and $1$ output neurons, respectively. The first two of these layers have a LeakyReLU activation while the final layer has no activation. 
%

The generator $\mathcal{G}$'s network structure consists of a Bidirectional Long Short-Term Memory (BLSTM) \cite{s_e_BLSTM} with two LSTM layers with 200 neurons each. This is followed by two fully connected layers, the first with 300 output neurons and a LeakyReLU \cite{Maas13rectifiernonlinearities} activation and the second 257 output neurons with a 'Learnable' Sigmoid activation function. 
This Learnable Sigmoid is given as:
\begin{equation}
\label{eq:5}
y_{\mathrm{learnable-sigmoid}} = \frac{\beta}{1+e^{-\alpha x}}
\end{equation}
where $\beta$ is a hyper-parameter (default to $1.2$) and $\alpha$ is a learnable parameter. In the original work the authors found that allowing $\beta$ to be learnable did not increase performance. 

\section{Proposed System - MetricGAN+/-}\label{mg-}
\subsection{MetricGAN+/- Framework}


The framework proposed in this work, MetricGAN+/-, expands on MetricGAN+ in one major way - we introduce an additional network, a `de-generator' $\mathcal{N}$ which, given an input signal $x$, will attempt to output a signal with a non-perfect score of metric $Q'$. The key idea of this extension is to allow $\mathcal{D}$ to observe a wider range of metrics scores outside of those present in the training data. The output audio of $\mathcal{N}$'s mask $\mathbf{M}_\mathcal{N}$ applied to noisy magnitude spectogram $|\mathbf{X}|$ is defined as $y$ and its feature space representation as $\mathbf{Y}_f$. An extra term is appended to the objective function of $\mathcal{D}$ that accounts for the prediction of the $Q'$ scores of these `de-enhanced' signals:
\begin{multline}
\label{eq:3}
\hspace*{-1.5ex}
L_{D,\mathrm{MG+/-}} 
= \mathbb{E}[(\mathcal{D}(\mathbf{S}_f,\mathbf{S}_f) -1)^2
+ (\mathcal{D}(\mathbf{\hat{S}}_f,\mathbf{S}_f)- Q'(\hat{s},s))^2\\
+ (\mathcal{D}(\mathbf{X}_f,\mathbf{S}_f)- Q'(x,s))^2
+(\mathcal{D}(\mathbf{Y}_f,\mathbf{S}_f) -Q'(y,s))^2]
\end{multline}
where $y$ represents the output of the de-generator network on the noisy signal $x$.
The objective function of $\mathcal{N}$ is given as
\begin{equation}\label{eq:4}
L_{\mathcal{N},\mathrm{MG+/-}} = \mathbb{E}[(\mathcal{D}(\mathbf{Y}_f,\mathbf{S}_f) -w)^2],\,\,\,  \mathrm{for}\,0 < w < 1,
\end{equation}
where $w$ is a hyper-parameter corresponding to the value of $Q'$ we train $\mathcal{N}$ to output signals with. The objective function of $\mathcal{G}$ is the same as for MetricGAN+, as given in (\ref{eq:2}).
The training of $\mathcal{N}$ is the same as the training of $\mathcal{G}$ depicted in Fig.~\ref{fig:gen_train} except that $\mathcal{G}$ is replaced by $\mathcal{N}$, $\hat{s}$, $\mathbf{\hat{S}}_f$ by $y$, $\mathbf{Y}_f$ and the $1$ in the \ac{MSE} by $w$.
This means that the training of $\mathcal{N}$ is influenced entirely by its performance as assessed by $\mathcal{D}$, in the same manner as  $\mathcal{G}$. 
%
We use an identical network structure to $\mathcal{G}$ for $\mathcal{N}$ 
- We leave to future work to change this structure, as well as related hyper-parameters such as the clamp threshold. 

%
The training of MetricGAN+/- is  similar to that that of MetricGAN+ given above with slight differences. Firstly $\mathcal{D}$ is trained using (\ref{eq:3}); as a result the replay buffer now contains both enhanced and de-enhanced data, effectively doubling its size. After $\mathcal{D}$'s training, $\mathcal{N}$ is trained using (\ref{eq:4}). Then $\mathcal{G}$ is trained as usual.

\section{Experiments}\label{sect:Results}
%
\subsection{Dataset}
%
The dataset used in the following experiments is VoiceBank-DEMAND \cite{vb-demand}. This
is a popular and commonly used dataset for single channel speech enhancement.
Its training set consists of $11572$ clean $s[n]$ and noisy $x[n]$ speech audio file pairs, mixed at 4  \acp{SNR} {$0$, $5$, $10$, $15$}~dB. Eight noise files are sourced from the DEMAND~\cite{demand} noise dataset - a cafeteria, a car interior, a kitchen, a meeting, a metro station, a restaurant, a train station and heavy traffic, and two others a babble noise and a speech-shaped noise. 
The utterances in the set vary in length from around $10$ seconds to $1$. The training set contains speech from 28 different speakers ($14$ male, $14$ female), English or Scottish accents. The testset containing $824$ utterances is mixed at SNRs of {$2.5$, $7.5$, $12.5$ and $17.5$}~dB, with five different noises which do not appear in the training set from the DEMAND corpus (bus, cafe, office, public square and living room) and contains speech from two (one male, one female) speakers who do not appear in the training set.


In order to better assess the system's ability to generalise to unseen noise types and recording scenarios as well as real recordings, we also assess performance of the models trained on VoiceBank-DEMAND training set on the test set of the CHiME3~\cite{barker_asru2015} challenge dataset. This test set consists of $1320$ real and $1320$ simulated noisy clean/speech pairs. For the real recordings the clean `reference' is a close-talk headset microphone which may also capture some of the background noise from the recording environment. There are $6$ channels of noisy recordings; we select the $5$th channel as input to the single channel systems as it has the most direct energy to the speaker and is the one used in the baseline system of the CHiME3 challenge. The recording environments of the real data and background noise of the simulated are a bus,  a cafe, a pedestrian area and a street junction. The simulated data is not mixed at any fixed \ac{SNR}, instead an ideal mixing SNR is calculated from analysis of the clean reference and the background recording.
%
\subsection{Experiment Setup}
The aim of the following experiments is to compare the performance of the baseline system MetricGAN+ which is available as part of the SpeechBrain \cite{speechbrain} toolkit with our  extension, MetricGAN+/-.
The Adam optimiser~\cite{kingma2017adam} with a learning rate of $0.0005$ is used. The \ac{STFT} is used with a DFT length of {$L_{\mathrm{DFT}}=$}$512$, a window length of $512$ ($32$~ms) at sampling frequency of $f_s=16$~kHz and a hop (overlap) length $256$ ($16$~ms), resulting in a $50\%$ overlap between frames. The minimum value in the time frequency masks $\mathbf{M}_\mathcal{G}$ and $\mathbf{M}_\mathcal{N}$ is set to $\xi=0.05$.

We experiment with both PESQ and STOI as objective $Q$ and different values of $w$. The values of $w$ are selected such that they correspond to sparely populated values of $Q'$ in the dataset. We also performed one experiment (denoted by * in Table~\ref{tab:1}) where  the value of $\beta$ in $\mathcal{N}$'s Learnable Sigmoid activation as given in (\ref{eq:5}) to also learned (in addition to $\alpha$).  Additionally, we experiment with reducing the size of the replay buffer training step for $\mathcal{D}$, via modifying $H$. In order to ensure that our performance gain does not come entirely from the larger $H$ in MetricGAN+/-, we report also the baseline MetricGAN+ performance with $H$ set to $0.4$. 
\subsection{Experiment Results}
%
Table \ref{tab:1} shows the performance of MetricGAN+/- relative to the MetricGAN+ baseline and the unprocessed noisy audio on the VoiceBank-DEMAND testset. We also compare performance with a second baseline system
SEGAN  \cite{DBLP:conf/interspeech/PascualBS17}, a state-of-the-art speech enhancement system. For more comparison baseline performances the interested reader is referred to Table 3 in \cite{fu21_interspeech}, which shows that MetricGAN+ with a PESQ objective outperforms all systems listed terms of  PESQ score. We assess this performance using \ac{PESQ} and \ac{STOI} and also using the Composite \cite{composite} Measure, where Csig, Cbak and Covl are intrusive measures of speech signal quality, background noise reduction quality, and overall quality respectively.

\begin{table}[!htb]
\centering
\caption{Performance of MetricGAN+ (MG+) and MetricGAN+/- (MG+/-) on VoiceBank-DEMAND test set for objective PESQ (P) or STOI (S), * denotes the simulation where $\beta$ is made learnable} \label{tab:1}
\resizebox{\columnwidth}{!}{%
\begin{tabular}{lccccccccc}
\rowcolor[HTML]{C0C0C0}
\textbf{Model Name} & \textbf{Obj.} & \textbf{w}   & \textbf{H}         & \textbf{P} & \textbf{S} & \multicolumn{1}{l}{\textbf{Csig}} & \multicolumn{1}{l}{\textbf{Cbak}} & \textbf{Covl} \\ \hline
\cellcolor[HTML]{C0C0C0}\textit{Noisy}             & \textit{-} & \textit{-}     & \multicolumn{1}{l|}{\textit{-}} & \textit{1.97} & \textit{92}   & \textit{3.35}                     & \textit{2.44}                     & \textit{2.63} \\
\cellcolor[HTML]{C0C0C0}MG+ (P)~\cite{fu21_interspeech}   & P                       & -              & \multicolumn{1}{l|}{0.2}          & 3.05          & 93            & 4.03                     & 2.87                     & 3.52 \\
\cellcolor[HTML]{C0C0C0}MG+ (S)   & S                        & -              & \multicolumn{1}{l|}{0.2}          & 2.42          & \textbf{93.4} & 3.56                              & 2.58                              & 2.97          \\
\cellcolor[HTML]{C0C0C0}SEGAN~\cite{DBLP:conf/interspeech/PascualBS17}                       & -            & -            & \multicolumn{1}{l|}{-}          & 2.42         & 92.5           & 3.61                             & 2.61                             & 3.01          \\\hline
\cellcolor[HTML]{C0C0C0}MG+  & P                        & -              & \multicolumn{1}{l|}{0.4}          & 3.17          & 92.3            & 4.05                     & 2.91                     & 3.59 \\

\cellcolor[HTML]{C0C0C0}MG+/-          & P               & 1.0                      & \multicolumn{1}{l|}{0.2}          & 3.20           & 93.0            & 4.08                              & 2.94                              & 3.62          \\
\cellcolor[HTML]{C0C0C0}MG+/-          & P               & 0.50                     & \multicolumn{1}{l|}{0.2}          & \textbf{3.22} & 91.3          & 4.05                              & 2.94                              & 3.62          \\
\cellcolor[HTML]{C0C0C0}MG+/-          & P               & 0.45                    & \multicolumn{1}{l|}{0.2}          & 3.21          & 91.9          & 4.09                              & \textbf{2.95}                              & 3.64          \\
\cellcolor[HTML]{C0C0C0}MG+/-*          & P               & 0.45                    & \multicolumn{1}{l|}{0.2}          & 3.17          & 93.0          & \textbf{4.16}                              & 2.93                              & \textbf{3.65}          \\
\cellcolor[HTML]{C0C0C0}MG+/-          & P               & 0.45                   & \multicolumn{1}{l|}{0.1}          & 3.13          & 92.1          & 4.05                              & 2.91                              & 3.58          \\
\cellcolor[HTML]{C0C0C0}MG+/-          & P               & 0.30                    & \multicolumn{1}{l|}{0.2}          & 3.04          & 93.0            & 4.07                              & 2.88                              & 3.55          \\
\cellcolor[HTML]{C0C0C0}MG+/-          & S               & 0.45                    & \multicolumn{1}{l|}{0.1}          & 2.13          & 93.2          & 3.04                              & 2.42                              & 2.56 
    \\
\cellcolor[HTML]{C0C0C0}MG+/-          & S               & 0.30                    & \multicolumn{1}{l|}{0.2}          & 2.31          & 93.3          & 3.19                              & 2.49                              & 2.72          
        
\end{tabular}
}
\end{table}

The first  four rows  in Table~\ref{tab:1} present the results the un-enhanced noisy data and of different baselines. The results for the baseline MetricGAN+ models shown here are obtained using the implementation in SpeechBrain \cite{speechbrain}. 
Further simulations are conducted for various values of  hyperparameter $w$ used in the training of $\mathcal{N}$.
Table~\ref{tab:1} shows a clear improvement in PESQ score for PESQ objective MetricGAN+/- models over the baseline MetricGAN+ ($3.05$  vs $3.22$ PESQ), and also versus the  PESQ value reported in \cite{fu21_interspeech} of $3.15$. We also observe increase in the composite measure scores. Interestingly, there is an improvement even when $w = 1$, which is the case where $\mathcal{N}$ and $\mathcal{G}$ have the same objective, and thus $\mathcal{N}$ also learns to enhance. We hypothesise that this is due to slight variations in the outputs of $\mathcal{N}$ and $\mathcal{G}$ during training, as well as the increased replay buffer size compared to the baseline. Highest performance in terms of PESQ score is obtained with a $w$ value set to 0.5, which means that $\mathcal{N}$ attempts to produce signals with a PESQ score of 3. We speculate that this performance increase is due to there being few clean/noisy pairs in the training set with a PESQ score around this value.

By making the $\beta$ parameter in $\mathcal{N}$'s activation function learnable, we observe a slight improvement against the baseline, as well as increased Csig and Covl scores versus all other simulations. 
We find also that increasing $H$ in the baseline MetricGAN+ from $0.2$ to $0.4$ such that its size is comparable to MetricGAN+/-'s does slightly improve PESQ score. This is contrary to the findings in \cite{fu21_interspeech} where they report no improvement for values larger than $0.2$. Larger values of $H$ will drastically increase the training time requirement of the system. We speculate that a better understanding of what $\mathcal{D}$ leans from the replay buffer training and better curation of its contents is the key to further performance gains, as well as reduced training time required.

\subsection{Spectogram Analysis}
\begin{figure}[!ht]
    \centering
    \def\svgwidth{\columnwidth} 
 	\graphicspath{{figures/}}  
 	\resizebox{.9\columnwidth}{!}{
    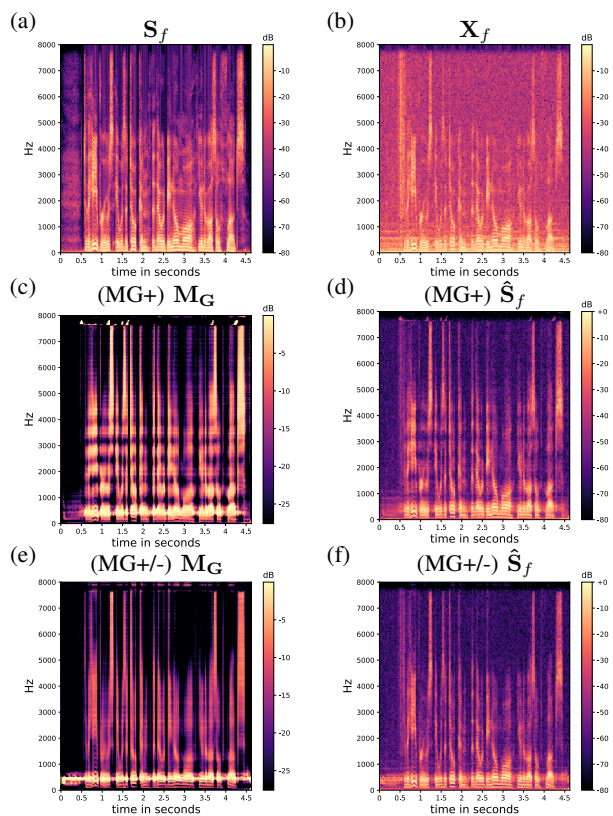 
    }
    \caption{Spectograms of: (a) clean reference features $\mathbf{S}_f$, (b) noisy features $\mathbf{X}_f$, (c) Mask $\mathbf{M}_\mathcal{G}$ and (d) enhanced output
    $\hat{S}_f$ for MetricGAN+ baseline PESQ objective model, (e) Mask $\mathbf{M}_\mathcal{G}$ and (f) enhanced output $\hat{S}_f$ for MetricGAN+/-  PESQ objective model. Source audio file is \texttt{p232\_014.wav} of VoiceBank-DEMAND testset.}
    \label{fig:pesq_specs}
\end{figure}

Fig.~\ref{fig:pesq_specs} shows the spectrograms of the clean reference $|\mathbf{S}|$, noisy input $|\mathbf{X}|$, generator output mask $\mathbf{M}_\mathcal{G}$ and this mask applied $|\mathbf{\hat{S}}|$ for baseline MetricGAN+, MetricGAN+/- ($w=0.45$) PESQ objective models. The mask in Fig.~\ref{fig:pesq_specs}~(e) attempts to remove low frequency signal content while boosting the area corresponding to the frequency curve of the fundamental speech frequencies. Futhermore, the baseline MetricGAN+ PESQ model in Fig.~\ref{fig:pesq_specs}~(c, d) attenuates the signal in the initial non speech region, while the MetricGAN+/- model in Fig.~\ref{fig:pesq_specs}~(e, f) suppresses less energy around $400$~Hz over the whole utterance. 
This artefact can already be observed in the baseline MetricGAN+ but is more prominent for the proposed method, which could explains the relatively low Cbak score for this method. 
We hypothesise that this is a results of $\mathcal{D}$ not learning to properly penalise errors in this region, perhaps due to the additional influence of $\mathcal{N}$'s outputs on it's training.
Further experimentation is required to fully understand this artefact. 
\subsection{Generalisation To Unseen Data}
\begin{table}[!ht]
\caption{Performance on real component of CHiME3 test set}
\centering
\resizebox{0.75\columnwidth}{!}{%
\begin{tabular}{llllll}
\rowcolor[HTML]{C0C0C0}
\textbf{Model Type} & \textbf{PESQ} & \textbf{STOI} & \textbf{Csig} & \textbf{Cbak} & \textbf{Covl} \\ \hline
\cellcolor[HTML]{C0C0C0}\textit{Noisy} & \textit{1.37} & \textit{44.0} & \textit{\textbf{2.96}} & \textit{1.42} & \textit{2.09} \\
\cellcolor[HTML]{C0C0C0}MG+ PESQ & 1.54 & 45.8 & 2.67 & 2.09 & 2.00 \\
\cellcolor[HTML]{C0C0C0}MG+ STOI & 1.24 & 44.7 & 2.45 & 1.84 & 1.76 \\ \hline
\cellcolor[HTML]{C0C0C0}MG+/- PESQ & \textbf{1.76} & 44.3 & 2.86 & \textbf{2.03} & \textbf{2.20} \\
\cellcolor[HTML]{C0C0C0}MG+/- STOI & 1.22 & \textbf{45.3} & 2.31 & 1.81 & 1.67
\end{tabular}
}
\label{tab:chime-results-real}
\end{table}
\vspace{-0.5cm}
\begin{table}[!ht]
\caption{Performance on simulated component of CHiME3 test set}
\centering
\resizebox{0.75\columnwidth}{!}{%
\begin{tabular}{lccccc}
\rowcolor[HTML]{C0C0C0}
\textbf{Model Type} & \textbf{PESQ} & \textbf{STOI} & \textbf{Csig} & \textbf{Cbak} & \textbf{Covl} \\ \hline
\cellcolor[HTML]{C0C0C0}\textit{Noisy} & \textit{1.27} & \textit{87.0} & \textit{2.61} & \textit{1.39} & \textit{1.88} \\
\cellcolor[HTML]{C0C0C0}MG+ PESQ & 2.14 & 87.4 & 3.05 & 2.31 & 2.53 \\
\cellcolor[HTML]{C0C0C0}MG+ STOI & 1.52 & \textbf{88.9} & 2.75 & 2.07 & 2.08 \\ \hline
\cellcolor[HTML]{C0C0C0}MG+/- PESQ & \textbf{2.38} & 86.1 & \textbf{3.17} & \textbf{2.41} & \textbf{2.70} \\
\cellcolor[HTML]{C0C0C0}MG+/- STOI & 1.47 & 88.5 & 2.62 & 2.02 & 1.99
\end{tabular}
}
\label{tab:chime-results-simu}
\end{table}



Tables~\ref{tab:chime-results-real} and~\ref{tab:chime-results-simu} shows the performance of the baseline MetricGAN+ and the best performing proposed MetricGAN+/- systems on this test set. 
We observe an increased performance in terms of PESQ, Csig, Cbak and Covl between PESQ objective MetricGAN+/- and the baseline, as well as a slight improvement in STOI score for STOI objective MetricGAN+/-. This suggests that $\mathcal{D}$'s access to the de-generated signals produced by $\mathcal{N}$ allows $\mathcal{G}$ in MetricGAN+/- systems to generalise better to unseen environments. 
\section{Conclusion}\label{sect: Conclusion}
In this work, we present an extension to the MetricGAN+ baseline framework, which improves its performance in terms of \ac{PESQ} score and related measures, as well as improving its generalisation to unseen data. We find that training the discriminator network on a wider range of metric scores and with a larger replay buffer achieves greater performance that the baseline system.

\bibliographystyle{IEEEtran}
\bibliography{test}

\begin{thebibliography}{10}
\providecommand{\url}[1]{#1}
\csname url@samestyle\endcsname
\providecommand{\newblock}{\relax}
\providecommand{\bibinfo}[2]{#2}
\providecommand{\BIBentrySTDinterwordspacing}{\spaceskip=0pt\relax}
\providecommand{\BIBentryALTinterwordstretchfactor}{4}
\providecommand{\BIBentryALTinterwordspacing}{\spaceskip=\fontdimen2\font plus
\BIBentryALTinterwordstretchfactor\fontdimen3\font minus
  \fontdimen4\font\relax}
\providecommand{\BIBforeignlanguage}[2]{{%
\expandafter\ifx\csname l@#1\endcsname\relax
\typeout{** WARNING: IEEEtran.bst: No hyphenation pattern has been}%
\typeout{** loaded for the language `#1'. Using the pattern for}%
\typeout{** the default language instead.}%
\else
\language=\csname l@#1\endcsname
\fi
#2}}
\providecommand{\BIBdecl}{\relax}
\BIBdecl

\bibitem{FFASRHaebUmbach}
R.~Haeb-Umbach, J.~Heymann, L.~Drude, S.~Watanabe, M.~Delcroix, and
  T.~Nakatani, ``Far-field automatic speech recognition,'' \emph{Proceedings of
  the IEEE}, vol. 109, no.~2, pp. 124--148, 2021.

\bibitem{XMM+15}
F.~Xiong, B.~Meyer, N.~Moritz, R.~Rehr, J.~Anem{\"{u}}ller, T.~Gerkmann,
  S.~Doclo, and S.~Goetze, ``Front-end technologies for robust {ASR} in
  reverberant environments - spectral enhancement-based dereverberation and
  auditory modulation filterbank features,'' \emph{EURASIP Journal on Advances
  in Signal Processing}, vol. 2015, no.~1, 2015.

\bibitem{barker_asru2015}
J.~Barker, R.~Marxer, E.~Vincent, and S.~Watanabe, ``The third {CH}i{ME} speech
  separation and recognition challenge: dataset, task and baselines,'' in
  \emph{Proc. IEEE Automatic Speech Recognition and Understanding Workshop
  (ASRU) 2015}, Scottsdale, Arizona, USA, 2015, pp. 504--511.

\bibitem{reddy2021interspeech}
C.~K. Reddy, H.~Dubey, K.~Koishida, A.~Nair, V.~Gopal, R.~Cutler, S.~Braun,
  H.~Gamper, R.~Aichner, and S.~Srinivasan, ``Interspeech 2021 deep noise
  suppression challenge,'' in \emph{INTERSPEECH}, 2021.

\bibitem{RGH22_AttTASNET}
W.~Ravenscroft, S.~Goetze, and T.~Hain, ``{Att-TasNet: Attending to Encodings
  in Time-Domain Audio Speech Separation of Noisy, Reverberant Speech
  Mixtures},'' \emph{Frontiers in Signal Processing}, vol.~2, 2022.

\bibitem{dolcoMVDRenhance}
M.~Tammen and S.~Doclo, ``Deep multi-frame {MVDR} filtering for
  single-microphone speech enhancement,'' in \emph{ICASSP 2021 - 2021 IEEE
  International Conference on Acoustics, Speech and Signal Processing (ICASSP),
  Toronto, Canada}, 2021, pp. 8443--8447.

\bibitem{Moritz2017}
N.~Moritz, K.~Adilo{\u{g}}lu, J.~Anem{\"{u}}ller, S.~Goetze, and B.~Kollmeier,
  ``Multi-channel speech enhancement and amplitude modulation analysis for
  noise robust automatic speech recognition,'' \emph{Computer Speech \&
  Language}, vol.~46, pp. 558--573, Nov 2017.

\bibitem{GWK+14}
S.~Goetze, A.~Warzybok, I.~Kodrasi, J.~Jungmann, B.~Cauchi, J.~Rennies,
  E.~Habets, A.~Mertins, T.~Gerkmann, S.~Doclo, and B.~Kollmeier, ``{A Study on
  Speech Quality and Speech Intelligibility Measures for Quality Assessment of
  Single-Channel Dereverberation Algorithms},'' in \emph{Proc. Int. Workshop on
  Acoustic Signal Enhancement (IWAENC 2014)}, Antibes, France, Sep. 2014.

\bibitem{bagchi2018spectral}
D.~Bagchi, P.~Plantinga, A.~Stiff, and E.~Fosler-Lussier, ``Spectral feature
  mapping with mimic loss for robust speech recognition,'' 2018.

\bibitem{Chai2018AcousticsguidedE}
L.~Chai, J.~Du, and C.-H. Lee, ``Acoustics-guided evaluation (age): a new
  measure for estimating performance of speech enhancement algorithms for
  robust asr,'' \emph{ArXiv}, vol. abs/1811.11517, 2018.

\bibitem{stoi_loss}
S.-W. Fu, T.-W. Wang, Y.~Tsao, X.~Lu, and H.~Kawai, ``End-to-end waveform
  utterance enhancement for direct evaluation metrics optimization by fully
  convolutional neural networks,'' \emph{IEEE/ACM Transactions on Audio,
  Speech, and Language Processing}, vol.~26, no.~9, 2018.

\bibitem{fu2018qualitynet}
S.-W. Fu, Y.~Tsao, H.-T. Hwang, and H.-m. Wang, ``Quality-net: An end-to-end
  non-intrusive speech quality assessment model based on blstm,'' in \emph{Proc
  of Interspeech 2018}, 09 2018, pp. 1873--1877.

\bibitem{PESQNet}
T.~F. Ziyi~Xu, Maximilian~Strake, ``Deep noise suppression with non-intrusive
  {PESQNet} supervision enabling the use of real training data,'' in
  \emph{Proc. of INTERSPEECH}, Brno, Czechia, 2021, pp. 1--5.

\bibitem{PESQNet2}
Z.~Xu, M.~Strake, and T.~Fingscheidt, ``{Deep Noise Suppression Maximizing
  Non-Differentiable PESQ Mediated by a Non-Intrusive PESQNet},''
  \emph{IEEE/ACM Transactions on Audio, Speech, and Language Processing},
  vol.~30, pp. 1572--1585, 2022.

\bibitem{fu21_interspeech}
S.-W. Fu, C.~Yu, T.-A. Hsieh, P.~Plantinga, M.~Ravanelli, X.~Lu, and Y.~Tsao,
  ``{MetricGAN+: An Improved Version of MetricGAN for Speech Enhancement},'' in
  \emph{Proc. Interspeech 2021}, 2021, pp. 201--205.

\bibitem{FuLTL19}
S.-W. Fu, C.-F. Liao, Y.~Tsao, and S.~de~Lin, ``Metricgan: Generative
  adversarial networks based black-box metric scores optimization for speech
  enhancement,'' in \emph{Proceedings of the 36th International Conference on
  Machine Learning, ICML 2019, 9-15 June 2019, Long Beach, California, USA},
  2019, pp. 2031--2041.

\bibitem{pesq}
A.~Rix, J.~Beerends, M.~Hollier, and A.~Hekstra, ``Perceptual evaluation of
  speech quality ({PESQ})-a new method for speech quality assessment of
  telephone networks and codecs,'' in \emph{2001 IEEE International Conference
  on Acoustics, Speech, and Signal Processing. Proceedings}, 2001.

\bibitem{stoi}
C.~H. Taal, R.~C. Hendriks, R.~Heusdens, and J.~Jensen, ``An algorithm for
  intelligibility prediction of time–frequency weighted noisy speech,''
  \emph{IEEE Transactions on Audio, Speech, and Language Processing}, vol.~19,
  no.~7, pp. 2125--2136, 2011.

\bibitem{reddy2021dnsmos}
C.~Reddy, V.~Gopal, and R.~Cutler, ``{DNSMOS: A Non-Intrusive Perceptual
  Objective Speech Quality metric to evaluate Noise Suppressors},'' in
  \emph{ICASSP 2020}, October 2020, pp. 6493--6497.

\bibitem{s_e_BLSTM}
F.~Weninger, H.~Erdogan, S.~Watanabe, E.~Vincent, J.~Le~Roux, J.~R. Hershey,
  and B.~Schuller, ``Speech enhancement with lstm recurrent neural networks and
  its application to noise-robust asr,'' in \emph{Latent Variable Analysis and
  Signal Separation}, E.~Vincent, A.~Yeredor, Z.~Koldovsk{\'y}, and
  P.~Tichavsk{\'y}, Eds., 2015.

\bibitem{Maas13rectifiernonlinearities}
A.~L. Maas, A.~Y. Hannun, and A.~Y. Ng, ``Rectifier nonlinearities improve
  neural network acoustic models,'' in \emph{in ICML Workshop on Deep Learning
  for Audio, Speech and Language Processing}, 2013.

\bibitem{vb-demand}
C.~{Valentini-Botinhao}, ``Noisy speech database for training speech
  enhancement algorithms and tts models,'' University of Edinburgh. Centre for
  Speech Technology Research (CSTR), 2017, doi: 10.7488/ds/2117.

\bibitem{demand}
J.~Thiemann, N.~Ito, and E.~Vincent, ``{DEMAND: a collection of multi-channel
  recordings of acoustic noise in diverse environments},'' Jun. 2013,
  {Supported by Inria under the Associate Team Program VERSAMUS}.

\bibitem{speechbrain}
M.~Ravanelli, T.~Parcollet, P.~Plantinga, A.~Rouhe, S.~Cornell, L.~Lugosch,
  C.~Subakan, N.~Dawalatabad, A.~Heba, J.~Zhong, J.-C. Chou, S.-L. Yeh, S.-W.
  Fu, C.-F. Liao, E.~Rastorgueva, F.~Grondin, W.~Aris, H.~Na, Y.~Gao, R.~D.
  Mori, and Y.~Bengio, ``Speechbrain: A general-purpose speech toolkit,'' 2021.

\bibitem{kingma2017adam}
D.~Kingma and J.~Ba, ``Adam: A method for stochastic optimization,''
  \emph{International Conference on Learning Representations}, 12 2014.

\bibitem{DBLP:conf/interspeech/PascualBS17}
S.~Pascual, A.~Bonafonte, and J.~Serr{\`{a}}, ``{SEGAN:} speech enhancement
  generative adversarial network,'' in \emph{Proc. Interspeech}, 2017.

\bibitem{composite}
Z.~Lin, L.~Zhou, and X.~Qiu, ``A composite objective measure on subjective
  evaluation of speech enhancement algorithms,'' \emph{Applied Acoustics}, vol.
  145, pp. 144--148, 02 2019.

\end{thebibliography}



\end{document}